# Solution processable and optically switchable 1D photonic structures

Giuseppe M. Paternò[1], Chiara Iseppon[2], Alessia D'Altri[2], Carlo Fasanotti[2], Giulia Merati[2], Mattia Randi[2], Andrea Desii[1], Eva A. A. Pogna[3], Daniele Viola[3], Giulio Cerullo[3], Francesco Scotognella[3], Ilka Kriegel[4,5]*

[1] Center for Nano Science and Technology@PoliMi, Istituto Italiano di Tecnologia, Via Giovanni Pascoli, 70/3, 20133, Milan, Italy

[2] Dipartimento di Chimica, Materiali e Ingegneria Chimica "Giulio Natta", Politecnico di Milano, Piazza Leonardo da Vinci 32, 20133 Milano, Italy

[3] Dipartimento di Fisica, Politecnico di Milano, Piazza Leonardo da Vinci 32, 20133 Milano, Italy

[4] Department of Nanochemistry, Istituto Italiano di Tecnologia (IIT), via Morego, 30, 16163 Genova, Genova, Italy

[5] The Molecular Foundry, Lawrence Berkeley National Laboratory, Berkeley, California 94720, United States

*Corresponding author: ilka.kriegel@iit.it

**Abstract**
In this work, we report the first demonstration of a solution processable, optically switchable 1D photonic crystal by implementing phototunable doped metal oxide nanocrystals. The resulting device structure shows bi-photonic response with the photonic bandgap covering the visible spectral range and the plasmon resonance of the doped metal oxide the near infrared. By means of a facile photodoping process, we tuned the plasmonic response and switched effectively the optical properties of the photonic crystal, translating the effect from the near infrared to the visible. The ultrafast bandgap pumping induces a signal change in the region of the photonic stopband, with recovery times of several picoseconds, providing a step toward the ultrafast optical switching. Optical modeling uncovers the importance to understand largely the variations of the dielectric function of the photodoped material, and variations in the high frequency region of the Drude response are responsible for the strong switching in the visible after photodoping. Our device configuration offers unprecedented tunablility due to flexibility in device design, cover wavelength ranges from the visible to the near infrared. Our findings indicate a new protocol to modify the optical response of photonic devices by optical triggers only.

**Introduction**
Solution processed photonic crystals facilitate the option of easy device fabrication on various substrates, but it further brings along the option of introducing material with diverse and complementary features. For example, the integration of plasmonic nanoparticles into a photonic crystal gives a new degree of freedom in the design of its optical response.[1–3] Recently, the interest of the scientific community went towards the study of heavily doped semiconductor nanoparticles that show plasmonic response in the near infrared (NIR). Such materials are very sensitive to changes in their dielectric function induced by variations in their carrier density through doping control.[4–8] This, in turn, is directly related to their Drude response and ultimately resulting in a blue shift of their near infrared plasmon resonance. Notably, in recent years additional routes have been opened to add reversibility to this effect by introducing and extracting extra carriers via capacitive charging, a process that permits to switch the optical response of the material in the NIR by modulating the amount of charges injected/extracted via an applied bias.[9–14] Furthermore, one can easily introduce extra carriers in such materials via optical triggers, i.e. photodoping.[9,10,15–18]

. In such process, photons with energies larger than the optical band gap promote electrons to the conduction band, leaving holes in the valence band. In the presence of a hole scavenger (i.e. chemical scavenger) these latter are extracted, leaving the system behind with additional electrons in their conduction band.

The integration of heavily doped semiconductor nanocrystals in one dimensional (1D) photonic crystals has been shown previously, for example by Puzzo et al. that reported silicon dioxide ($SiO_2$) /antimonium tin oxide (ATO) and ATO/ titanium dioxide ($TiO_2$) 1D photonic crystals.[19,20] However, the exploitation of their tunable plasmonic response has not been demonstrated yet. This is particularly interesting as the carrier density modulation results in the active manipulation of their frequency dependent dielectric constant (the square of the refractive index) and the refractive index contrast in turn is what determines the photonic bandgap of the device. Thus, the modulation of the near infrared plasmonic response of the nanoparticle film can be employed to modulate the overall optical response of the photonic system, with the effect being two-fold: a change in the near infrared plasmonic response,[21] as well as a variation of the photonic band gap.[22–25]

In this work we studied, for the first time, such plasmonic/photonic effect in a photonic crystal composed of alternating nanoparticle layers of $SiO_2$ and indium tin oxide (ITO). The latter is a very prominent heavily doped semiconductor with carrier densities in the range of $10^{21} cm^{-3}$ in which the reversible tunability of its near infrared plasmonic response by various means has been demonstrated.[16,26–28] The photonic device covers the entire spectrum from the visible (photonic band gap) to the near infrared (free carrier response in ITO), supported by optical modeling. We show the switching of its optical response via photodoping in steady state and ultrafast time scales, together taking a step towards a solution processed, contactless, all-optical switching device.

**Methods**

*Nanoparticles*: $SiO_2$ nanoparticles were purchased from Sigma-Aldrich (Ludox SM-30) and it was diluted with distilled water to a final concentration of 5 wt. %. The size of the nanoparticles is 8 nm. ITO nanoparticles were purchased from Gentech Nanomaterials and and it was diluted with distilled water to a final concentration of 5 wt. %. The size of the nanoparticles is 20-30 nm. The dispersions were sonicated for 60 minutes at room temperature and filtered with a 0.45 μm PVDF filter.

*Fabrication of the 1D photonic crystal*: A glass substrate was washed in isopropanol and then in acetone in a sonicating bath for 5 minutes. Then, organic contaminants were removed from the substrate via an oxygen plasma treatment.

The photonic crystal has been fabricated using a spin coater Laurell WS-400- 6NPP-Lite. The rotation speeds for the deposition were 2000 rotations per minute (rpm). After each deposition, the sample has been annealed for 10 minutes at 350 °C on a hot plate under the fume hood.

*Scanning electron microscope characterization*: the microscope was a Tescan MIRA3. The measurement has been performed at a voltage of 5 kV and backscattered electrons have been detected. The sample has been covered with carbon paste to improve conductivity.

*Spectroscopic measurements*: The transmission spectrum of the photonic crystal has been measured with a Perkin Elmer spectrophotometer Lambda 1050 WB. The ultrafast spectroscopy measurement has been performed with a laser system based on a Ti-Sapphire chirp pulse amplified source (Coherent Lybra), with a maximum output energy of about 800 μJ, 1 kHz repetition rate, and central wavelength at about 780 nm. The pulse duration is about 150 fs. Excitation pulses at 260 nm were obtained via the sum frequency of the fundamental one at 780 nm and its second harmonic in a β-Barium borate (BBO) crystal. Pump pulses were focused in a 200 μm diameter spot. Probing was achieved in the visible by using white light generated in a thin sapphire plate. Spectra of chirp-free transient normalized transmission change ($\Delta T/T$) were collected by using an optical multichannel analyser (OMA) with a dechirping algorithm.

*CW Photodoping experiment*: after a UV irradiation of the sample with a LED at 310 nm for 5 minutes, an absorption spectrum has been acquired. The UV irradiation and the absorption measurement have been performed at room temperature in air.

*Transfer Matrix Method*: The transmission spectra of the photonic crystal were calculated by employing the transfer matrix method, widely used for the transcription of the optical response of multilayers.[29-34] The amplitude of the electric and magnetic fields are determines after the light wave propagates through the multilayer stacks considering the Maxwell equations with the proper boundary conditions. The matrix product of the characteristic transmission matrix through each layer gives the overall transmission, as it calculates the electric and magnetic fields at the output of the photonic structure. The assumption of isotropic, non-magnetic systems is valid for most dielectric materials and a normal incidence angle was implemented. The following system of output amplitudes has been solved:

$$\begin{bmatrix} E_0 \\ H_0 \end{bmatrix} = M_1 \cdot M_2 \cdot ... \cdot M_m \begin{bmatrix} E_m \\ H_m \end{bmatrix} = \begin{bmatrix} m_{11} & m_{12} \\ m_{21} & m_{22} \end{bmatrix} \begin{bmatrix} E_m \\ H_m \end{bmatrix} \qquad (1)$$

Here, $E_0$ and $H_0$ depict the amplitudes of electric and magnetic fields at the input, while $E_m$ and $H_m$ are the amplitudes at the output. The matrices for each layer $j$ are then given by:

$$M_j = \begin{bmatrix} A_j & B_j \\ C_j & D_j \end{bmatrix}, \qquad (2)$$

with $j=(1,2,...,m)$ and the elements of the transmission matrix *ABCD*:

$$A_j = D_j = \cos(\phi_j), B_j = -\left(\frac{i}{p_j}\right)\sin(\phi_j), C_j = -ip_j \sin(\phi_j). \qquad (3)$$

$n_j$ and $d_j$ are determined through the phase variation $\phi_j$, and depict the effective refractive index and the thickness of the layer $j$, respectively. For normal incidence the phase variation in the $j$-fold layer simplifies to $\phi_j = \left(\frac{2\pi}{\lambda}\right) n_j d_j$. The coefficient $p_j = \sqrt{\frac{\varepsilon_j}{\mu_j}}$ for the transverse electric wave and $q_j = 1/p_j$ for a transverse magnetic wave.

The transmission $t$ is then determined by:

$$t = \frac{2p_s}{(m_{11}+m_{12}p_0)p_s + (m_{11}+m_{12}p_0)} \qquad (4)$$

And the final light transmission given as:

$$T = \frac{p_0}{p_s}|t|^2 \qquad (5)$$

where $p_s$ represents the substrate and $p_0$ air.

**Results and Discussion**

Photonic crystals have been fabricated by solution processing of alternating layers from dispersion of SiO$_2$ and ITO nanoparticles, respectively. In Figure 1a we report the SEM cross section of the SiO$_2$/ITO 1D photonic crystal made of 5 bilayers. The relatively darker regions correspond to the SiO$_2$ layers, while the brighter ones to the ITO layers due to the lower atomic weight of silicon with respect to indium and tin. The average layer thickness for SiO$_2$ is 120 nm, while the one for ITO is 60 nm. In the upper panel of Figure 1b we show the absorption spectrum of the fabricated SiO$_2$/ITO photonic crystal, displaying a peak at around 500 nm and intense absorption in the near infrared. The peak in the visible can be ascribed to the photonic band gap of the photonic crystal as a result of the refractive index contrast $\Delta n$ between the alternating nanoparticle layers. The tail from about 1500 nm towards longer wavelengths instead arises from the plasmonic response of the ITO nanoparticles, brought about by the free carrier density $N$ in the range of $10^{20}$-$10^{21}$ cm$^{-3}$.[16,26–28] We remark here that for simplicity sake we give the optical response of the system in absorbance, rather than transmittance, although the visible peak of the photonic bandgap is not actually due to absorption. Absorbance however, takes into account all processes occurring in the sample due to absorption and reflection.

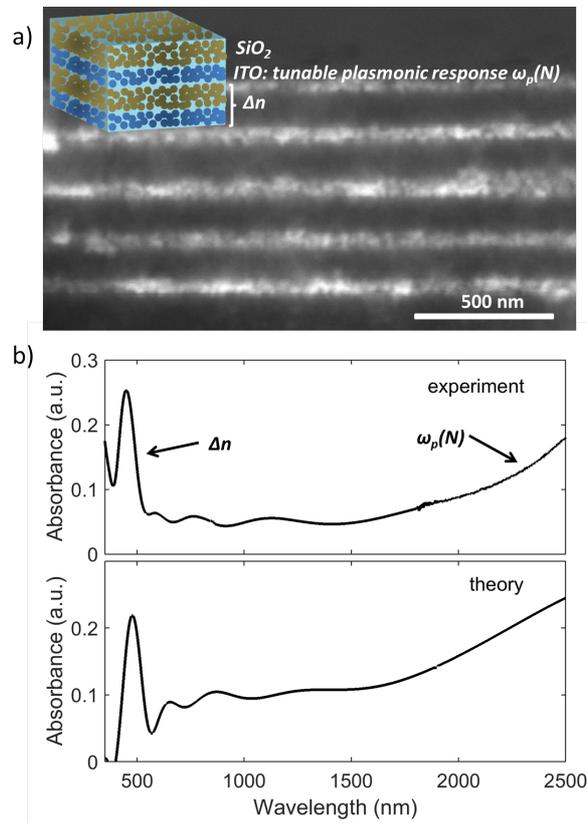

**Figure 1.** a) SEM cross section of the SiO$_2$/ITO 1D photonic crystal. The bright layers correspond to the higher contrast ITO nanoparticle layer, the darker layers to SiO$_2$. Inset: sketch of the SiO$_2$/ITO 1D photonic crystal indicating the contribution to the optical response: $\Delta n$ corresponds to the refractive index contrast, which is responsible for the photonic bandgap in the visible, while the high absorption in the near infrared is due to free carriers (i.e. plasmonic response) of the ITO nanoparticles depicted by the plasma frequency as a function of the carrier density $N$ ($\omega_p(N)$). b) Experimental (upper panel) and theoretical (lower panel) absorption spectrum of the 1D photonic structure.

We modeled the optical response of the photonic crystal by implementing two different models that account for the respective spectral contributions: firstly, we implement the transfer matrix theory, which takes into account the alternating refractive index of the periodic structure, and secondly we integrated the Drude model and the Maxwell-Garnet effective medium approximation (MG-EMA) to consider the contribution of the ITO nanoparticle films to the dielectric response of the photonic crystal.[35-37] The latter theory (MG-EMA) is implemented to account for a dense distribution of particles dispersed in a dielectric matrix, considering that the layer of ITO actually consists of ITO nanoparticles and air voids. The effective dielectric function is then given as:

$$\varepsilon_{eff,ITO}(\omega) = \varepsilon_{Air} \frac{2(1-\delta_{ITO})\varepsilon_{Air} - (1+2\delta_{ITO})\varepsilon_{ITO}(\omega)}{2(2+\delta_{ITO})\varepsilon_{Air} + (1+\delta_{ITO})\varepsilon_{ITO}(\omega)}, \quad (6)$$

where $\delta_{ITO}$ is the so called fill factor, depicting the fraction of the volume of the inclusions occupying the volume of the film, while $\varepsilon_{ITO}(\omega)$ is the frequency dependent Drude dielectric function of ITO and $\varepsilon_{Air} = (1.00059)^2$. Consequently, we obtain the effective refractive index by

$$n^2_{eff,ITO}(\omega) = \varepsilon_{eff,ITO}(\omega). \quad (7)$$

The dielectric function of ITO is described by the Drude theory, which takes into account the contribution from the free carriers to the optical response and is given by:

$$\varepsilon_{ITO}(\omega) = \varepsilon_1(\omega) + i\varepsilon_2(\omega), \quad (8)$$

where the real part is given as

$$\varepsilon_1(\omega) = \varepsilon_\infty - \frac{\omega_P^2}{(\omega^2 - \Gamma^2)}, \quad (9)$$

and the imaginary part as

$$\varepsilon_2(\omega) = \frac{\omega_P^2 \Gamma}{\omega(\omega^2 - \Gamma^2)}, \quad (10)$$

where $\omega$ is the frequency and $\Gamma$ is the damping (inverse of the carrier relaxation time). The plasma frequency $\omega_P$ is given as

$$\omega_P = \sqrt{\frac{Ne^2}{m^* \varepsilon_0}}, \quad (11)$$

where $N$ is the carrier density, $e$ is the electron charge, $m^*$ is the charge effective mass and $\varepsilon_0$ is the vacuum permittivity. This relation shows clearly that the plasmonic response is proportional to the square root of the carrier density ($\omega_p(N)$).

We also implemented the effective refractive index of the $SiO_2$ nanoparticle layer for the calculation of the alternating structure with the transfer matrix method. For details on the transfer matrix method please refer to the methods section.[32-34] In the lower panel of Figure 1b is given the theoretically obtained absorption spectrum, showing remarkable overlap with the experiment. Note that we used a film thickness of 110nm for the $SiO_2$ layer to match the position of the photonic bandgap. We obtained the best fit by employing a fill factor for the ITO nanoparticle film of $\delta_{ITO} = 0.3$. We used a significantly higher fill factor (0.95) for the $SiO_2$ films due to the much denser film, as observed in the SEM image in Figure 1a. We extracted a plasma frequency of $\omega_P$ = 9000 cm$^{-1}$ and from the latter determined the carrier density of 3.7·10$^{20}$ cm$^{-3}$, which is in agreement with common ITO samples.[16,26–28]

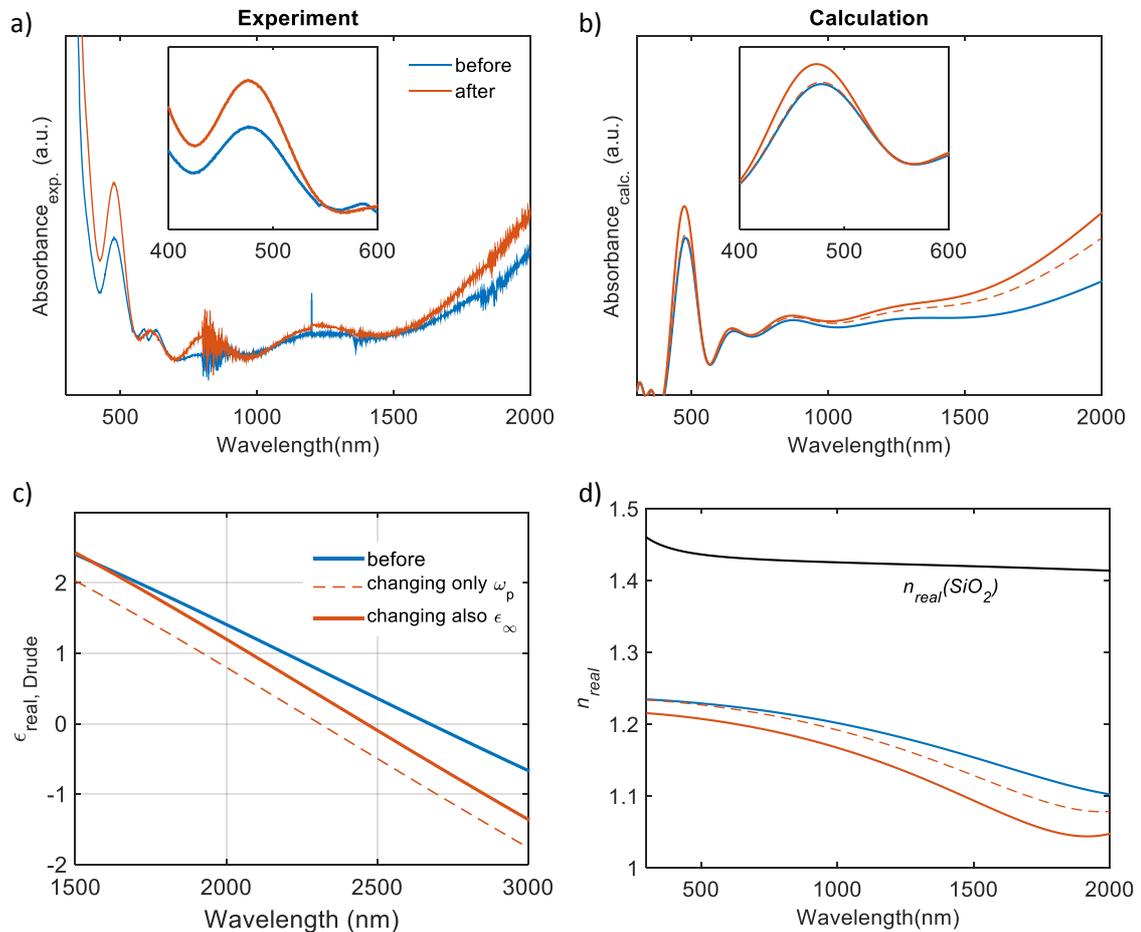

**Figure 2.** (a) Experimental absorption spectrum of the SiO$_2$/ITO photonic crystal, before and after UV exposure (photodoping); (b) calculation of the absorption spectrum with the transfer matrix method, integrated with the Drude model and the Maxwell-Garnet effective medium approximation; (c) real part of the Drude dielectric function of the ITO nanoparticles; (d) real part of the refractive index of the ITO nanocomposite film as employed in the calculation. Black curve: SiO$_2$, blue to red curves: ITO nanocomposite.

Having assigned the spectral signature of our device structure, we performed a photodoping experiment by exposing ITO/SiO$_2$ photonic structure to UV light (310nm LED) for five minutes and measured its absorbance before and after illumination (Figure 2). We observe an increase of the absorption in the NIR regime after illumination due to the enhanced carrier density related to $\omega_p(N)$ (red curve in Figure 2a) Notably, we also observe a significant effect on the region of the photonic bandgap (see inset to Figure 2a), related to the difference in the refractive index of the two alternating layers. To extract a better understanding of these spectral changes we modelled this system also after photodoping, and the results of this simulation are given in Figure 2b-d. The blue curve gives the same modelling result of the photonic device as shown in Figure 1b (lower panel). Figure 2c shows the Drude dielectric function of the ITO nanoparticles, showing the cross-over from positive to negative values in the so called epsilon near zero regime.[38,39] After UV light exposure, in our model we artificially increase the $\omega_p(N)$, illustrating the increased carrier density after photodoping. From this simulation (Figure 2b, dashed curve) you can see that we were able to reproduce the variation observed in the NIR by adjusting this parameter, indicating an important role of the enhanced carrier density on the optical spectra. However, this simple picture does not permit to reproduce the shift observed in the region of the photonic bandgap. When looking at the real part of the refractive index $n_{real}$ of our nanoparticle film, as given in Figure 2d it becomes clear that the refractive index contrast $\Delta n$ is barely altered in the visible spectral range (red dashed curves versus blue curves, respectively) when increasing $\omega_p(N)$. Interestingly, after inducing a notable difference in the $\Delta n$ after varying further the high frequency dielectric constant ($\varepsilon_\infty$), we are able to explain the experimental changes of the photonic bandgap in the visible spectral range (Figure 2b), as this largely alters the refractive index contrast with $SiO_2$ (black curve in Figure 2d). Actually, it has been shown that increased doping influences also the region of the interband transitions in relation to the Moss-Burstein shift of the bandgap or, in other words, the additional Pauli blocking due to the occupation of additional levels in the conduction band.[40] The sample becomes more transparent in the high frequency region which is expressed as an increase in the high frequency dielectric constant ($\varepsilon_\infty$). Our results highlight the paramount importance to carefully study the variation of the dielectric function upon photodoping of such materials. In particular, despite a major role has been given to the increase in charge carrier density that dictates the optical modulation of charged metal oxide structures,[41] our results prove unambiguously that also the high frequency dielectric constant is markedly altered by the photodoping process. Indeed, when looking at the Drude response of the varied dielectric functions after photodoping, one can see a strong influence in particular on the epsilon near zero regime of the real part of the Drude dielectric function (Figure 2c), highlighting the importance to study in detail the variations of the dielectric response to understand the effects of carrier injection upon photodoping. Moreover, our results illustrate how the all optical control over the epsilon near zero regime can be employed to design photo-switchable devices.

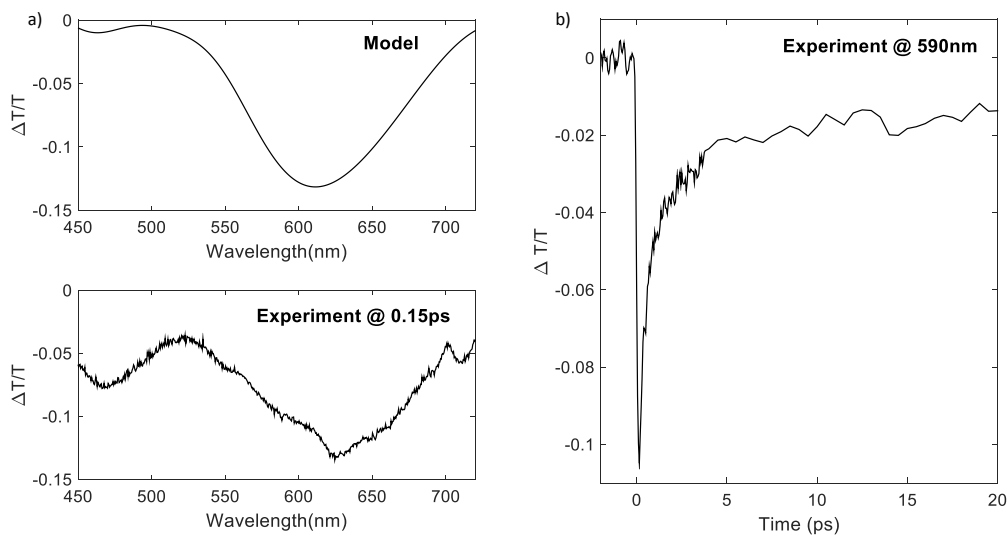

**Figure 3.** a) Transient transmission spectrum at 0.15 ps after photoexcitation by pumping the 1D photonic crystal at 260 nm (4.77 eV). Upper panel displays the model and lower panel displays the experiment. b) Differential transmission dynamics at 590 nm.

It has been shown in previous works that the photodoping can also be detected on ultrafast time scales in the picosecond time frame. For instance, in fluorine and indium doped cadmium oxide (FICO) nanocrystals an ultrafast modulation of the plasmon resonance has been shown after band edge excitation due to the ultrafast photodoping.[42] The authors extracted a monoexponential recovery time within several picoseconds, indicative of the recombination of electron and hole. These measurements thus detect the temporarily increased carrier density after photodoping as long as the electron is placed in the conduction band and before recombining with the excited state hole. Similar results have been observed also in indium cadmium oxide nanocrystals[38] and in indium tin oxide nanopillars.[26] In a similar way we also performed an ultrafast photodoping experiment on our photonic crystal device by pumping the sample with a 150 fs laser pulse at 260 nm (4.77 eV), well above the band gap of ITO (about 4 eV). We show the results in Figure 3 showing the differential transmission spectra of the $SiO_2$ – ITO 1D photonic crystal at the maximum ΔT/T at around 150 fs and the differential transmission dynamic at 590 nm close to the maximum photonic bandgap. Note that the specifications of the photonic structure were not exactly the same as in the previous example and therefore the photonic bandgap shifted to higher wavelengths. Remarkably, we observe an ultrafast recovery time of the signal within about 20 picoseconds, and a change of the transmission of approximately 10% at moderate pump intensities (0.5 mJ/cm$^2$). In addition, we show here that we can simulate the differential transmission spectrum by employing the extracted parameters from the steady state model, as used above and implementing the transmission changes from the transfer matrix method calculated as:

$$\frac{\Delta T}{T} = \frac{T_{ON} - T_{OFF}}{T_{OFF}} \qquad (12)$$

where $T_{ON}$ accounts for the transmission of the white light through the sample after photo-excitation at 260 nm. We obtained a good agreement between the experimental and simulated spectra. We remark here that reminding to the previous reports on the ultrafast photodoping of FICO nanocrystals, in both cases a change of the high frequency dielectric constant has been considered to describe the ultrafast response. Although for FICO nanocrystals a positive variation of approximately 0.4 was observed, we see here a negative variation of the same magnitude. The difference in sign between both experiments might well be due to the different systems, in which once single nanocrystals in a dilute solution have been considered, while we study densely packed films. Additionally, also difference in the material band structure might play a role. To clarify this point further studies of the ultrafast response of ITO nanocrystals also in the near infrared are required. Nevertheless, our results show that ultrafast photodoping of ITO nanocrystals translate to the transient signal change of the photonic bandgap in the visible spectral range.

**Conclusion**
In this paper we report the solution processed fabrication of photo-switchable 1D photonic crystals with bi-photonic effects: the photonic stop band in the visible range and the plasmonic resonance in the near infrared. The incorporation of heavily doped semiconductor nanoparticle layers adds an additional degree of tunability to the device due to capacitive charge injection triggered by photodoping of the ITO nanoparticles. The all-optical switching is obtained by exposing the device to UV light which results in a modification of both, the photonic bandgap and the plasmon resonance. Optical modeling revealed that the photodoping has an effect not only on the plasma frequency $\omega_p(N)$ that depends on the carrier density, but also on the high frequency refractive index, most probably due to the Moss-Burstein effect of heavily doped semiconductors. This highlights that elevated optical modeling is required to extract fundamental variations in the dielectric function of photo-switchable nanoparticles. Ultrafast switching times of the photonic bandgap in the visible have been extracted after transient photodoping, suggesting our system for ultrafast optical signal processing with high transmission changes and recovery time in the picosecond time frame. Taken together, we propose a new optically switchable system, in which the optical response can be varied in the steady state and ultrafast times in a contactless manner.

**Acknowledgements**

This project has received funding from the European Union's Horizon 2020 research and innovation programme (MOPTOPus) under the Marie Skłodowska-Curie grant agreement No. [705444], as well as (SONAR) grant agreement no. [734690] and H2020 ETN SYNCHRONICS under grant agreement 643238. Work at the Molecular Foundry was supported by the Office of Science, Office of Basic Energy Sciences, of the U.S. Department of Energy under Contract No. DE-AC02-05CH11231.

**References**
(1) Morandi, V.; Marabelli, F.; Amendola, V.; Meneghetti, M.; Comoretto, D. Colloidal Photonic Crystals Doped with Gold Nanoparticles: Spectroscopy and Optical Switching Properties. *Adv. Funct. Mater.* **2007**, *17*, 2779–2786.
(2) Aluicio-Sarduy, E.; Callegari, S.; Figueroa del Valle, D. G.; Desii, A.; Kriegel, I.; Scotognella, F. Electric Field Induced Structural Colour Tuning of a Silver/titanium Dioxide Nanoparticle One-Dimensional Photonic Crystal. *Beilstein J. Nanotechnol.* **2016**, *7*, 1404–1410.
(3) Robbiano, V.; Giordano, M.; Martella, C.; Stasio, F. D.; Chiappe, D.; de Mongeot, F. B.; Comoretto, D. Hybrid Plasmonic–Photonic Nanostructures: Gold Nanocrescents Over Opals. *Adv. Opt. Mater.* **2013**, *1*, 389–396.
(4) Luther, J. M.; Jain, P. K.; Ewers, T.; Alivisatos, A. P. Localized Surface Plasmon Resonances Arising from Free Carriers in Doped Quantum Dots. *Nat. Mater.* **2011**, *10*, 361–366.
(5) Dorfs, D.; Härtling, T.; Miszta, K.; Bigall, N. C.; Kim, M. R.; Genovese, A.; Falqui, A.; Povia, M.; Manna, L. Reversible Tunability of the Near-Infrared Valence Band Plasmon Resonance in $Cu_{2-x}Se$ Nanocrystals. *J. Am. Chem. Soc.* **2011**, *133*, 11175–11180.
(6) Scotognella, F.; Della Valle, G.; Srimath Kandada, A. R.; Dorfs, D.; Zavelani-Rossi, M.; Conforti, M.; Miszta, K.; Comin, A.; Korobchevskaya, K.; Lanzani, G.; *et al.* Plasmon Dynamics in Colloidal $Cu_{2-x}Se$ Nanocrystals. *Nano Lett.* **2011**, *11*, 4711–4717.
(7) Kriegel, I.; Jiang, C.; Rodríguez-Fernández, J.; Schaller, R. D.; Talapin, D. V.; da Como, E.; Feldmann, J. Tuning the Excitonic and Plasmonic Properties of Copper Chalcogenide Nanocrystals. *J. Am. Chem. Soc.* **2012**, *134*, 1583–1590.
(8) Della Valle, G.; Scotognella, F.; Kandada, A. R. S.; Zavelani-Rossi, M.; Li, H.; Conforti, M.; Longhi, S.; Manna, L.; Lanzani, G.; Tassone, F. Ultrafast Optical Mapping of Nonlinear Plasmon Dynamics in $Cu_{2-x}Se$ Nanoparticles. *J. Phys. Chem. Lett.* **2013**, *4*, 3337–3344.
(9) Kriegel, I.; Scotognella, F.; Manna, L. Plasmonic Doped Semiconductor Nanocrystals: Properties, Fabrication, Applications and Perspectives. *Phys. Rep.* **2017**, *674*, 1–52.
(10) Agrawal, A.; Johns, R. W.; Milliron, D. J. Control of Localized Surface Plasmon Resonances in Metal Oxide Nanocrystals. *Annu. Rev. Mater. Res.* **2017**, *47*, 1–31.
(11) Llordes, A.; Runnerstrom, E. L.; Lounis, S. D.; Milliron, D. J. Plasmonic Electrochromism of Metal Oxide Nanocrystals. In *Electrochromic Materials and Devices*; Mortimer, R. J.; Rosseinsky, D. R.; Monk, P. M. S., Eds.; Wiley-VCH Verlag GmbH & Co. KGaA, 2013; pp. 363–398.
(12) Lounis, S. D.; Runnerstrom, E. L.; Llordes, A.; Milliron, D. J. Defect Chemistry and Plasmon Physics of Colloidal Metal Oxide Nanocrystals. *J. Phys. Chem. Lett.* **2014**, *5*, 1564–1574.
(13) Wang, Y.; Runnerstrom, E. L.; Milliron, D. J. Switchable Materials for Smart Windows. *Annu. Rev. Chem. Biomol. Eng.* **2016**, *7*, 283–304.
(14) Runnerstrom, E. L.; Llordés, A.; Lounis, S. D.; Milliron, D. J. Nanostructured Electrochromic Smart Windows: Traditional Materials and NIR-Selective Plasmonic Nanocrystals. *Chem. Commun.* **2014**, *50*, 10555–10572.
(15) Schimpf, A. M.; Knowles, K. E.; Carroll, G. M.; Gamelin, D. R. Electronic Doping and Redox-Potential Tuning in Colloidal Semiconductor Nanocrystals. *Acc. Chem. Res.* **2015**, *48*, 1929–1937.
(16) Schimpf, A. M.; Lounis, S. D.; Runnerstrom, E. L.; Milliron, D. J.; Gamelin, D. R. Redox Chemistries and Plasmon Energies of Photodoped In2O3 and Sn-Doped In2O3 (ITO) Nanocrystals. *J. Am. Chem. Soc.* **2015**, *137*, 518–524.
(17) Schimpf, A. M.; Gunthardt, C. E.; Rinehart, J. D.; Mayer, J. M.; Gamelin, D. R. Controlling Carrier Densities in Photochemically Reduced Colloidal ZnO Nanocrystals: Size Dependence and Role of the Hole Quencher. *J. Am. Chem. Soc.* **2013**, *135*, 16569–16577.

(18) Schimpf, A. M.; Thakkar, N.; Gunthardt, C. E.; Masiello, D. J.; Gamelin, D. R. Charge-Tunable Quantum Plasmons in Colloidal Semiconductor Nanocrystals. *ACS Nano* **2014**, *8*, 1065–1072.
(19) Puzzo, D. P.; Bonifacio, L. D.; Oreopoulos, J.; Yip, C. M.; Manners, I.; Ozin, G. A. Color from Colorless Nanomaterials: Bragg Reflectors Made of Nanoparticles. *J. Mater. Chem.* **2009**, *19*, 3500–3506.
(20) Bonifacio, L. D.; Lotsch, B. V.; Puzzo, D. P.; Scotognella, F.; Ozin, G. A. Stacking the Nanochemistry Deck: Structural and Compositional Diversity in One-Dimensional Photonic Crystals. *Adv. Mater.* **2009**, *21*, 1641–1646.
(21) Willets, K. A.; Van Duyne, R. P. Localized Surface Plasmon Resonance Spectroscopy and Sensing. *Annu. Rev. Phys. Chem.* **2007**, *58*, 267–297.
(22) John, S. Strong Localization of Photons in Certain Disordered Dielectric Superlattices. *Phys. Rev. Lett.* **1987**, *58*, 2486–2489.
(23) Yablonovitch, E. Inhibited Spontaneous Emission in Solid-State Physics and Electronics. *Phys. Rev. Lett.* **1987**, *58*, 2059–2062.
(24) Joannopoulos, J. D. *Photonic Crystals: Molding the Flow of Light*; Princeton University Press: Princeton, 2008.
(25) Bellingeri, M.; Chiasera, A.; Kriegel, I.; Scotognella, F. Optical Properties of Periodic, Quasi-Periodic, and Disordered One-Dimensional Photonic Structures. *Opt. Mater.* **2017**, *72*, 403–421.
(26) Guo, P.; Schaller, R. D.; Ketterson, J. B.; Chang, R. P. H. Ultrafast Switching of Tunable Infrared Plasmons in Indium Tin Oxide Nanorod Arrays with Large Absolute Amplitude. *Nat. Photonics* **2016**, *10*, 267–273.
(27) Llordes, A.; Garcia, G.; Gazquez, J.; Milliron, D. J. Tunable near-Infrared and Visible-Light Transmittance in Nanocrystal-in-Glass Composites. *Nature* **2013**, *500*, 323–326.
(28) Garcia, G.; Buonsanti, R.; Runnerstrom, E. L.; Mendelsberg, R. J.; Llordes, A.; Anders, A.; Richardson, T. J.; Milliron, D. J. Dynamically Modulating the Surface Plasmon Resonance of Doped Semiconductor Nanocrystals. *Nano Lett.* **2011**, *11*, 4415–4420.
(29) Criante, L.; Scotognella, F. Low-Voltage Tuning in a Nanoparticle/Liquid Crystal Photonic Structure. *J. Phys. Chem. C* **2012**, *116*, 21572–21576.
(30) Guduru, S. S. K.; Kriegel, I.; Ramponi, R.; Scotognella, F. Plasmonic Heavily-Doped Semiconductor Nanocrystal Dielectrics: Making Static Photonic Crystals Dynamic. *J. Phys. Chem. C* **2015**, *119*, 2775–2782.
(31) Kriegel, I.; Scotognella, F. Tunable Light Filtering by a Bragg Mirror/heavily Doped Semiconducting Nanocrystal Composite. *Beilstein J. Nanotechnol.* **2015**, *6*, 193–200.
(32) Chiasera, A.; Scotognella, F.; Criante, L.; Varas, S.; Valle, G. D.; Ramponi, R.; Ferrari, M. Disorder in Photonic Structures Induced by Random Layer Thickness. *Sci. Adv. Mater.* **2015**, *7*, 1207–1212.
(33) Malitson, I. H. Interspecimen Comparison of the Refractive Index of Fused Silica. *J. Opt. Soc. Am.* **1965**, *55*, 1205–1208.
(34) Born, M.; Wolf, E. *Principles of Optics: Electromagnetic Theory of Propagation, Interference and Diffraction of Light*; Cambridge University Press, 2000.
(35) Li, S.-Y.; Niklasson, G. A.; Granqvist, C. G. Plasmon-Induced near-Infrared Electrochromism Based on Transparent Conducting Nanoparticles: Approximate Performance Limits. *Appl. Phys. Lett.* **2012**, *101*, 71903.
(36) Mendelsberg, R. J.; Garcia, G.; Li, H.; Manna, L.; Milliron, D. J. Understanding the Plasmon Resonance in Ensembles of Degenerately Doped Semiconductor Nanocrystals. *J. Phys. Chem. C* **2012**, *116*, 12226–12231.

(37) Mendelsberg, R. J.; Garcia, G.; Milliron, D. J. Extracting Reliable Electronic Properties from Transmission Spectra of Indium Tin Oxide Thin Films and Nanocrystal Films by Careful Application of the Drude Theory. *J. Appl. Phys.* **2012**, *111*, 63515.

(38) Diroll, B. T.; Guo, P.; Chang, R. P. H.; Schaller, R. D. Large Transient Optical Modulation of Epsilon-Near-Zero Colloidal Nanocrystals. *ACS Nano* **2016**, *10*, 10099–10105.

(39) Runnerstrom, E. L.; Kelley, K. P.; Sachet, E.; Shelton, C. T.; Maria, J.-P. Epsilon-near-Zero Modes and Surface Plasmon Resonance in Fluorine-Doped Cadmium Oxide Thin Films. *ACS Photonics* **2017**, *4*, 1885–1892.

(40) Fujiwara, H.; Kondo, M. Effects of Carrier Concentration on the Dielectric Function of ZnO:Ga and ${\mathrm{In}}_{2}{\mathrm{O}}_{3}:\mathrm{Sn}$ Studied by Spectroscopic Ellipsometry: Analysis of Free-Carrier and Band-Edge Absorption. *Phys. Rev. B* **2005**, *71*, 75109.

(41) Mendelsberg, R. J.; McBride, P. M.; Duong, J. T.; Bailey, M. J.; Llordes, A.; Milliron, D. J.; Helms, B. A. Dispersible Plasmonic Doped Metal Oxide Nanocrystal Sensors That Optically Track Redox Reactions in Aqueous Media with Single-Electron Sensitivity. *Adv. Opt. Mater.* **2015**, *3*, 1293–1300.

(42) Kriegel, I.; Urso, C.; Viola, D.; De Trizio, L.; Scotognella, F.; Cerullo, G.; Manna, L. Ultrafast Photodoping and Plasmon Dynamics in Fluorine–Indium Codoped Cadmium Oxide Nanocrystals for All-Optical Signal Manipulation at Optical Communication Wavelengths. *J. Phys. Chem. Lett.* **2016**, *7*, 3873–3881.